\documentstyle[anta, twoside]{article}

\input{psfig.sty}

\def\Journal#1#2#3#4{(#1) {#2} {\bf #3}, #4}

\def\AAp{\em Astron. Astrophys.}

\def\ApJ{\em Astrophys.~J.}

\def\MNRAS{\em Mon. Not. R.~Astron. Soc.}

\newcommand{\HI}{{\rm H\,\scriptstyle I}}

\begin{document}

\markboth{I. Yusifov}{Pulsars and the Warp of the Galaxy}

\thispagestyle{plain}
\setcounter{page}{165}

\title{Pulsars and the Warp of the Galaxy}

\author{I. Yusifov}

\address{Department of Astronomy \& Space Sciences,
              Faculty of Arts \& Sciences,\\
              Erciyes University, Talas Yolu, 38039 Kayseri, Turkey\\}

\maketitle

\abstract{This paper studies the asymmetries of pulsar distributions
relative to the Galactic plane in various Galactic longitudes. It is
shown that the observed asymmetric distribution may be explained by the
warped and flaring (i.e. increase of scale-height at the peripheries of
the Galaxy) structure of the Galaxy. At the peripheries of the Galaxy,
the amplitude of the warp, derived from these data, may be as high as
$\sim$1 kpc. The scale-height of the pulsar distribution increases
exponentially from $\sim$0.5 kpc to  $\sim$1 kpc while increasing the
galactocentric distance from 5 kpc to 15 kpc. The warp and flare
parameters derived from the pulsar data are compared with the stellar
and gaseous warp and flare parameters of the Galaxy. }

\section{Introduction}

The existence of a Galactic warp has been known since the first radio
surveys of the Galaxy (Burton 1988). Furthermore, it was shown that
the distribution of various Galactic components at the peripheries of
the Galaxy also shows a flaring structure; i.e., growth of
scale-height with increasing Galactic radius. Subsequently, the
warped and flaring structure of the Galaxy was studied for young
OB stars and old stellar components of the Galaxy, as well as for the
Galactic dust emission at $240\,\mu$m (e.g. Alard 2000; Drimmel et
al. 1999; Drimmel \& Sperge 2001; L\'opez-Corredoira et al. 2002a).

The stability and the precise shape and nature of the Galactic
warp, however, still remain unclear. The Large Magellanic Cloud, the
Sagittarius dwarf galaxy and intergalactic accretion are considered as
possible sources of generating Galactic warps (see for example
Garc\'{\i}a-Ruiz et al. 2002; Tsuchiya 2002; L\'opez-Corredoira et
al. 2002b; and Bailin 2003).

It is generally accepted that the progenitors of pulsars are young
OB stars, and if their Galactic distributions show warped structure
(Drimmel et al. 1999), then the distribution of pulsars also must
reveal a similar structure. As objects observable from large
distances, pulsars may be more appropriate to study the warped and
flaring structure of the Galaxy.

The recent high-frequency, sensitive Parkes Multibeam Pulsar
Surveys (PMPS) (Manchester et al. 2001; Morris et al. 2002; and Kramer
et al. 2003) revealed many more distant pulsars with high Dispersion
Measures (DM). These and all other available observational data of
pulsars are collected in new ATNF Pulsar Catalogue (Manchester et al.
2002). The warp and flare of the Galaxy may be estimated analyzing
the longitude and latitude distribution of pulsars from this catalogue.

However, warp and flare are more pronounced at distances larger
than 10~kpc from the Galactic center and from the Sun. Unfortunately,
at present, the number of observable pulsars at these distances
is rather small. Probably, more precise estimates of the warp and
flare parameters may be possible when a greater number of pulsars will
be detected at the peripheries of the Galaxy. However, preliminary
studies of warp and flare may be done with the data already available,
which is presented in this contribution. Furthermore, warped and
flaring structures of the Galactic plane may be used in future to
attain a more detailed modeling of the Galactic electron distribution
and more precise estimates of distances for some distant pulsars.

\section{Observed Data and Warped Disc Model of the Galaxy }

Yusifov \& K\"u\c c\"uk (this volume, YK2003 hereafter) have studied
the Galactic distribution and the luminosity function of pulsars on the
basis of new ATNF catalogue of 1412 pulsars (Manchester et al. 2002).
We have used here the results of this study to derive the warp
and flare parameters of the Galaxy.

\begin{figure}[t]
\centerline{
\psfig{figure=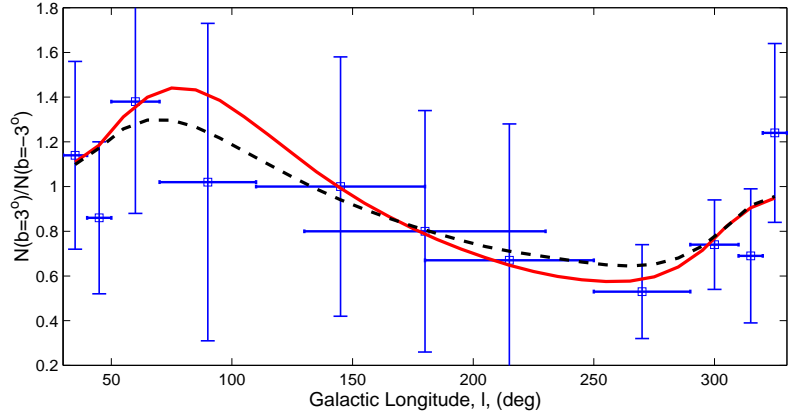,width=10.5truecm}
  }
\caption{Ratio of positive counts of pulsars to negative ones with
corresponding errors. Horizontal bars correspond to the width
of the considered longitude intervals. Dashed and solid lines
correspond to the best-fitting models with and without taking into
account data points at longitude $l=90\degr$.}
\label{Ratios}
\end{figure}

The estimates of the parameters of the warp and flare were made by the
method described in L\'opez-Corredoira et al. (2002a), with some
modifications. For this study we select ``normal'' pulsars at
Galactic latitudes $|l|\leq5\degr$, excluding binary and recycled
($\dot{P} <10^{-17}$ s/s), globular cluster and extragalactic pulsars.
We found the ratio of the cumulative number of pulsars above and
below the Galactic plane at various Galactic longitudes centered
at $b =\pm3\degr$; the results with the corresponding error bars
are plotted in Fig.~\ref{Ratios}. The errors in Fig.~\ref{Ratios}
are estimated assuming that the observed cumulative numbers follow the
Poisson distribution. In order to see the warping effect clearly,
pulsars located within $r\leq1$ kpc from the Sun were excluded from
consideration.

The number of pulsars at various longitudes varies dramatically, and
in order to gather a statistically significant number of pulsars,
as is seen from Fig.~\ref{Ratios}, the longitude width of the interval
was chosen to be variable. The latitude width of the intervals, which is
centered on $b=\pm3\degr$, was $\Delta b=4\degr$.

In spite of the existence of large error bars, the plot clearly
shows a sinusoidal behavior. If the warped structure were absent, the
ratio of counts should be nearly the same for all Galactic longitudes.
A similar behavior may be easily explained by the warp of the
Galactic plane.

The parameters of the warp will be found from Fig.~\ref{Ratios},
fitting the observed ratios with the modeled ones. The stellar statistic
equation has the form:
\begin{equation}
A(m)=\int^{\infty}_{0}r^2 D(r)\phi(M)dr\ ,
  \label{A(m)}
\end{equation}
where $A(m)$ is the number of stars per unit area of solid angle
$\omega$ at $m$ in interval d$m$, $\phi(\rm{M})$ is the luminosity
function and  $D(r)$  is the stellar density.  Converting  this
relation to the pulsar data for our purpose, we obtain:

\begin{equation}
N(l,b)=\omega \int^{r_1}_{r_2}N_0(l,b,r)\rho[R(l,b,r),z(l,b,r)]r^2 dr\ ,
 \label{N(l,b)}
\end{equation}
where $r$ and $R$ are the distances from the Sun and the Galactic
center, $\rho$ is the space density and

\begin{equation}
N_0(l,b,r)= \int^{L_{\rm max}}_{S_0r^2}\Phi(L)\frac{dL}{L}
 \label{N0(lbr)}
\end{equation}
is the local surface density of pulsars at distance $r$ from the
Sun. $S_0$ is the minimum detectable flux density, and $L_{\rm
max}=10^4$ mJy kpc$^2$ is the maximum luminosity of  pulsars at 1400
MHz. For the minimum detectable flux density we assume a derived value
of $S_0=0.07$ mJy from YK2003. The Galactic distribution and
luminosity function of pulsars accepted from YK2003 have the form:

\begin{equation}
\sigma(R)=C\left(\frac{R}{R_{\odot}}\right)^a
\exp\left(-b\frac{R-R_{\odot}}{R_{\odot}}\right)\ ,
 \label{sigmaR}
\end{equation}
where $\sigma(R)$ is the surface density of pulsars with parameters
$C=40$ kpc$^{-2}$, $a=1.12$, $b=3.2$, and $R_{\odot}=8.5$
kpc is Sun--Galactic center distance and

\begin{equation}
\Phi(L)={A_L \over \sigma_L \sqrt{2\pi}}
\exp{\biggl[-{1\over2}\biggl({{\rm log}L-{\rm log}L_O \over
\sigma_L}\biggr)^2 \biggr]}\ ,
\label{LF}
\end{equation}
where $A_L=67, \sigma_L=1.21$ and ${\rm log}\, L_O=-1.64$. The space
distribution then has the form:

\begin{equation}
\rho(l,b,r)=N_0(l,b,r) \left(\frac{R}{R_{\odot}}\right)^a
\exp\left(-b\frac{R-R_{\odot}}{R_{\odot}}\right)
\exp\left(-\frac{|z|}{h_z(R)}\right)
\frac{h_z(R)}{h_z(R_{\odot})}\ ,
 \label{Rolbr}
\end{equation}
where $h_z(R)$ is the scale-height of pulsar distributions. In order
to study flaring of pulsar distributions in the model is included
exponentially increasing scale-height by the relation:

 \begin{equation}
h_z(R)=h_z(R_{\odot})\exp\left(\frac{R-R_{\odot}}{h_{R,flare}}\right)\ ,
 \label{hzR}
\end{equation}
where $h_z(R_\odot)$ and $h_{R,flare}$ are free parameters. The
factor $\frac{h_z(R)}{h_z(R_{\odot})}$ in Eq. (6) is normalized
due to the variable scale-height of $h_z(R)$.

\begin{figure}[h]
\centerline{
\psfig{figure=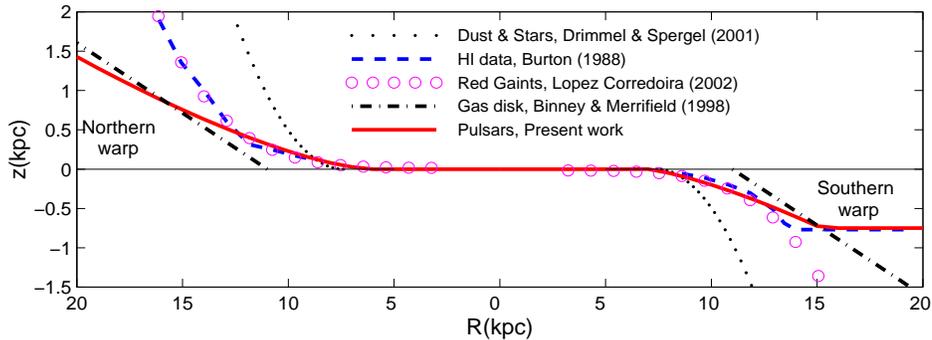,width=12.5truecm}
  }
\caption{Comparison of the maximum of amplitudes of the various
Galactic warps models, plotted as a function of galactocentric
distance. Data of all models were converted to $R_\odot=8.5$ kpc.}
\label{Warps}
\end{figure}

In order to take the warping structure into account in relation (6),
$|z|$ must be replaced by $|z-Z_w|$, where the function $Z_w(R,\phi)$
describes the vertical displacement of the warp. From the early
studies of $\HI$ it is known that the warping structure of the galaxy is
asymmetric (see for example Burton (1988)), and at the galactocentric
distances $R>14$ kpc the southern warp becomes constant with
height (see Fig.~\ref{Warps}). For this reason, for the warped model of
the Galactic plane, we include an additional free parameter, as the
galactocentric radius $R_{WS}$, from which the southern warp becomes
constant, and correspondingly $Z_w(R,\phi)$ is calculated by the
relation:

\begin{equation}
  Z_w(R,\phi)=\cases{C_W(R-R_W)^{b_w}\sin(\phi-\phi_W)+15, & for  $R\leq R_{WS}$\ , \cr
          & \cr
                C_W(R_{WS}-R_W)^{b_w}\sin(\phi-\phi_W)+15, & for  $R>R_{WS}$\ .  \cr}
  \label{ZwRfi}
\end{equation}
Here, $R_W$ is the galactocentric radius, from which the warp starts
(for $R < R_W$, $Z_w(R,\phi)=0$), $\phi$ is the galactocentric angle
taken in the direction of Galactic rotation with the Sun lying along
$\phi=0\degr$.

\section{Results and Discussion}

Using relations (2)--(8) we have fitted the observational data in
Fig.~\ref{Ratios} by the least mean square method and estimated the
warp and flare parameters in Eqs. (7) and (8). For the best-fitting
parameters we obtain:

\vspace{0.2cm}

\quad $h_z(R_\odot)$=580 pc, \quad  $h_{R,flare}$=14 kpc; \hfill           (9)

\noindent and

\quad $C_W$=37,\quad  $R_W$=6.5 kpc,\quad $b_w$=1.4,\quad
$\phi_w$=$14.5^\circ$\quad and\quad $R_{WS}$=15.2 kpc  \hfill                 (10)
\vspace{0.2cm}

\begin{figure}[t]
\centerline{
\psfig{figure=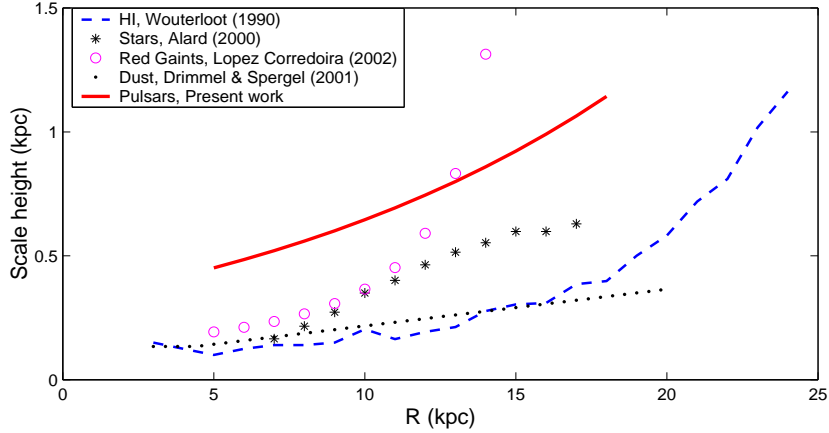,width=11truecm}
  }
\caption{Comparison of the scale-height variations of various Galactic
components plotted as a function of distance to the Galactic center.}
\label{Flare}
\end{figure}

The results of fitting by these parameters are shown by the solid
lines in Figs.~\ref{Ratios}--\ref{Flare}. In Figs.~\ref{Warps} and
\ref{Flare} we compared the warp and flare results of our study
with similar results of other studies. It is interesting to note
that the warp of the galactic disc derived from pulsar data is
more closely related to the gas disc warp (Binney \& Merrifield 1998;
Bailin 2003) than the stellar and dust warps, which are described in
L\'opez-Corredoira (2002a) and Drimmel \& Spergel (2001).

In fitting the observational data in Fig.~\ref{Ratios}, we used both
a symmetric and asymmetric warp. Calculations show that the
asymmetric warp better fits the observational data, and the obtained
southern asymmetry is very closely related to $\HI$ data (see
Fig.~\ref{Warps}).

We estimated the scale-height of the pulsar distribution
($h_z(R_\odot)=580$pc) here by quite a different method than that used
(e.g. Lyne et al. (1985) and references therein) and it occurred nearly
30\% greater than that of earlier estimates (450~pc). The large
scale-height of  pulsars may be caused by the additional fact of the
high-velocity origin of pulsars.

From Fig.~\ref{Flare} it is seen that, as for other components of the
Galaxy, the scale-height of the pulsar distribution also increased with
increasing galactocentric radius. This increase strongly deviates from
the red giant flaring and is nearly parallel to the $\HI$ and dust
flaring.

From previous studies it was known that the line of nodes or phase
angle of the warp ($\phi_w$) is $\approx 0\degr$ (see for example
Drimmel \& Spergel 2001) or even has a negative value
$\approx -5\degr$ (see for example  L\'opez-Corredoira et al. 2002).
But from fitting pulsar data in Fig.~1 we obtain $\phi_w \approx
15\degr$ for the phase angle of the warp.

\begin{figure}[t]
\centerline{
\psfig{figure=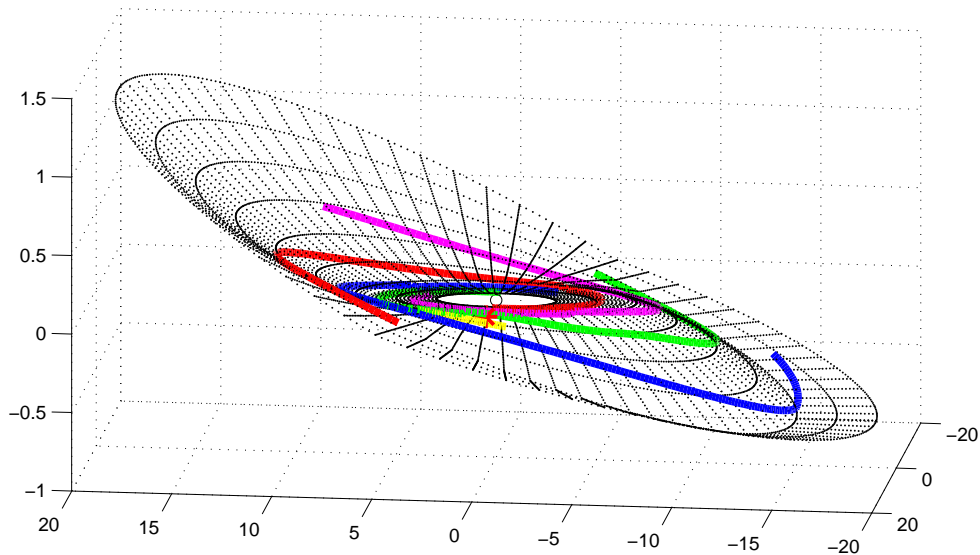,width=13truecm}
  }
\caption{A schematic 3D view of the Galactic warped plane together
with spiral arms from Cordes \& Lazio (2002). In order to present a
better appearance of the warping structure, the scale in the $z$
direction is considerably increased.}
\label{GPlane}
\end{figure}

Figure~\ref{GPlane} shows the general view of the Galactic plane
obtained in this study. There are very few pulsars at distances
$R>15$~kpc, and for this reason the precise values of warping and
flaring at these distances are still waiting to be elucidated.
Nevertheless, this warping structure of the Galactic plane may be very
helpful in distance estimates of some high-DM pulsars. The
inclusion of the warping and flaring structure of the Galaxy to the
large-scale modeling of the ISM electron distribution may serve as the
next step to improve the model.

\section*{Acknowledgments}

I would like to thank R.N. Manchester and the Parkes Multibeam Pulsar
Survey team for making the parameters of new pulsars available on the
internet prior to formal publication. I thank  Victor B. Cohen for help
in preparing the manuscript. This work has been partially supported  by
Erciyes University R/D project No. 01$-$052$-$1, Turkey. Extensive use
was made of both the Los Alamos preprint archive and the ADS system.

\section*{References}\noindent

\references

Alard, C. (2000) preprint [astro-ph/0007013].

Bailin, J. \Journal{2003}{\ApJ}{583}{L79}.  

Binney, J. \& Merrifield, M. (1998) {\em Galactic Astronomy},
Princeton University Press.

Burton, W.B. (1988) in {\em Galactic and Extragalactic Radio
 Astronomy},  eds. K.I. Kellermann \& G.L. Verschuur, Springer-Verlag,
 Berlin, p.~295.

Cordes, J.M. \& Lazio, T.J.W. (2002) preprint [astro-ph/0207156].

Drimmel, R., Smart, R.L. \& Lattanzi, M.G. \Journal{2000}{\AAp}{354}{67}.

Drimmel, R. \& Spergel, D.N. \Journal{2001}{\ApJ}{556}{181}. 

Garc\'{\i}a-Ruiz, I., Kuijken, K. \& Dubinski, J.
  \Journal{2002}{\MNRAS}{337}{459}. 

Kramer, M., Bell, J.F., Manchester, R.N., et al.
  \Journal{2003}{\MNRAS}{342}{1299}. 

L\'opez-Corredoira, M., Cabrera-Lavers, A., Garz\'on, F. \& Hammersley,
  P.L. (2002a) {\em Astron. Astrophys.} {\bf 394}, 883.

L\'opez-Corredoira, M., Betancort-Rijo, J. \& Beckman, J.E. (2002b)
  {\em Astron. Astrophys.} {\bf 386}, 169.

Lyne, A.G. , Manchester, R.N. \& Taylor, J.H. \Journal{1985}{\MNRAS}{213}{613}.

Manchester, R.N., et al. (2002) {\em ATNF Pulsar Catalogue},
          http://www.atnf.csiro.au/research/pulsar/psrcat

Manchester, R.N., Lyne, A.G., Camilo, F., et al. \Journal{2001}{\MNRAS}{328}{17}.

Morris, D.J., Hobbs, G., Lyne, A.G., et al. \Journal{2002}{\MNRAS}{335}{275}.

Tsuchiya, T. \Journal{2002}{\em New Astron.}{7}{293}. 

\end{document}